\documentclass[aps,report]{revtex4-1}

\usepackage{graphicx}
\usepackage{amssymb}
\usepackage{color}
\usepackage{dcolumn}
\usepackage{bm}

\newcommand{\figsizeone}{0.8}
\newcommand{\figsizetwo}{0.6}

\begin{document}

\draft
\title{Amplitude death in a ring of nonidentical nonlinear oscillators with unidirectional coupling}
\author{Jung-Wan Ryu}
\affiliation{Center for Theoretical Physics of Complex Systems, Institute for Basic Science, Daejeon 34051, South Korea}
\author{Jong-Ho Kim}
\affiliation{National Institute for Mathematical Sciences, Daejeon 34047, South Korea}
\author{Woo-Sik Son}
\affiliation{National Institute for Mathematical Sciences, Daejeon 34047, South Korea}
\affiliation{Center for Convergent Research of Emerging Virus Infection, Korea Research Institute of Chemical Technology, Daejeon 34114, South Korea}
\author{Dong-Uk Hwang}
\email{duhwang@nims.re.kr}
\affiliation{National Institute for Mathematical Sciences, Daejeon 34047, South Korea}

\begin{abstract}
We study the collective behaviors in a ring of coupled nonidentical nonlinear oscillators with unidirectional coupling,
of which natural frequencies are distributed in a random way.
We find the amplitude death phenomena in the case of unidirectional couplings
and discuss the differences between the cases of bidirectional and unidirectional couplings.
There are three main differences; there exists neither partial amplitude death nor local clustering behavior
but oblique line structure which represents directional signal flow on the spatio-temporal patterns in the unidirectional coupling case.
The unidirectional coupling has the advantage of easily obtaining global amplitude death in a ring of coupled oscillators with randomly distributed natural frequency.
Finally, we explain the results using the eigenvalue analysis of Jacobian matrix at the origin and
also discuss the transition of dynamical behavior coming from connection structure as coupling strength increases.
\end{abstract}

\maketitle
\narrowtext

\section{Introduction}

Coupled oscillators generate complex collective dynamics in a variety of fields such as biological oscillators \cite{Win80,Kur84,Str93},
Josephson junction arrays \cite{Had88a,Had88b,Bra95,Wie96}, neural networks \cite{Col95,Hop95}, and semiconductor lasers \cite{Var97,Hoh97}.
One of the important collective behaviors in coupled nonlinear oscillators is the amplitude death (AD) which refers to a situation
where individual oscillators cease to oscillate when the nonlinear dynamical systems are coupled \cite{Sax12}.
For the occurrence of AD in two diffusively coupled oscillators, a large mismatch of the natural frequencies of two oscillators is required \cite{Eli84,Mir90,Erm90,Aro90}.
If two identical oscillators are considered, the AD can be achieved by the existence of time delayed coupling \cite{Red98,Red99,Red00a,Red00b},
conjugate coupling \cite{Kar10}, dynamical coupling \cite{Kon03}, or nonlinear coupling \cite{Pra03,Pra10}.
The AD has also been studied in networks of coupled oscillators \cite{Erm90,Ata90,Rub00} and variety topologies
such as a ring \cite{Dod04,Kon04}, small world \cite{Hou03}, and scale free networks \cite{Liu09}.
Especially, oscillation suppressions in a ring of nonlinear oscillators have been studied in cases of different types of coupling. The AD can appear in a ring of oscillators with nonlinear \cite{Pra10} and delayed couplings \cite{Ata10, Gju14}. Oscillation death which has inhomogeneous steady state contrary to the homogeneous steady state of AD occurs in a ring of oscillator with non-local coupling \cite{Sch15}.

Most of the studies of AD in networks of coupled oscillators have focused on bidirectional (reciprocal) coupling cases.
However, there are many real systems with directional (non-reciprocal) coupling.
For instance, the coupling in neural networks with nearest-neighbor connections can be not only unequal but also of opposite sign
if one direction is excitatory and the other inhibitory \cite{Day01}.
It has been reported that the dynamics of collective neurons in these neural networks with directional couplings are different from bidirectional coupling cases \cite{Cha14, Ami16}.
Besides the applications of neuroscience, the directional coupling has been studied in terms of localizations in solid state physics.
In several literatures, it has been reported that the delocalization transition can be induced by an imaginary vector potential
in a disordered chain, which corresponds to the directional coupling \cite{Hat96, Efe97, Bro97}.
To the best of our knowledge, however, there have been few studies on the AD in the case of unidirectional coupling.
In the present work, we study the AD in a ring of coupled nonlinear oscillators with bidirectional and cyclically unidirectional couplings as two limiting cases,
of which natural frequencies are distributed in a random way.
We find the spatially-distributed AD phenomena in the case of unidirectional coupling and compare the results with well-known AD phenomena in the case of bidirectional coupling.
Finally, we explain the results using the eigenvalue analysis of Jacobian matrix at the origin,
and discuss the transition of dynamical behavior coming from connection structure as coupling strength increases.

This paper is organized as follows.
In Sec. II, a coupled Stuart-Landau limit cycle oscillators on a ring structures are introduced.
In Sec. III, we numerically obtain the spatio-temporal patterns in rings of coupled nonlinear oscillators with bidirectional and unidirectional couplings, respectively,
and classify non-amplitude death (NAD), partial amplitude death (PAD), and global amplitude death (GAD) regions. 
We discuss the differences between two coupling cases.
In Sec. IV, we explain the results of Sec. III using the eigenvalue analysis of Jacobian matrix at the origin.
Finally, we summarize our results in Sec. V.

\section{Models}

We use coupled Stuart-Landau limit cycle oscillators on a ring structure,
\begin{eqnarray}
\label{slosc}
 \dot{z}_j=(1+i \omega_j - |z_j|^2) z_j + k F(z_{j-1}, z_{j}, z_{j+1}),
\end{eqnarray}
where $j=1, \cdots, N$.
Here $F$ represents the diffusive coupling function between the nearest neighbors,
\begin{eqnarray}
\label{coupling}\nonumber
 F(z_{j-1}, z_{j}, z_{j+1}) &=& (z_{j-1} -2 z_{j} + z_{j+1})/2,\\\nonumber
 F(z_{j-1}, z_{j}, z_{j+1}) &=& (z_{j+1} - z_{j})
\end{eqnarray}
for bidirectional and cyclically unidirectional coupling cases, respectively.
$z_j$ are complex variables and $k$ is the coupling strength.
$\omega_j$ are the intrinsic angular frequency of the uncoupled $j$th limit cycle oscillator,
which is a random number between $\omega_{min}$ and $\omega_{max}$ with uniform probability distribution.
The ring structure has the periodic boundary condition, i.e., $z_{N+j}=z_{j}$.
Without coupling ($k=0$), $N$ oscillators which have unstable fixed points at origin exhibit the limit cycle with radii $1$ and angular frequencies $\omega_j$.
In the followings, we set $N=1000$.

\section{Spatio-temporal patterns}

\subsection{Bidirectional coupling case}

\begin{figure}
\begin{center}
\includegraphics[width=\figsizetwo\textwidth]{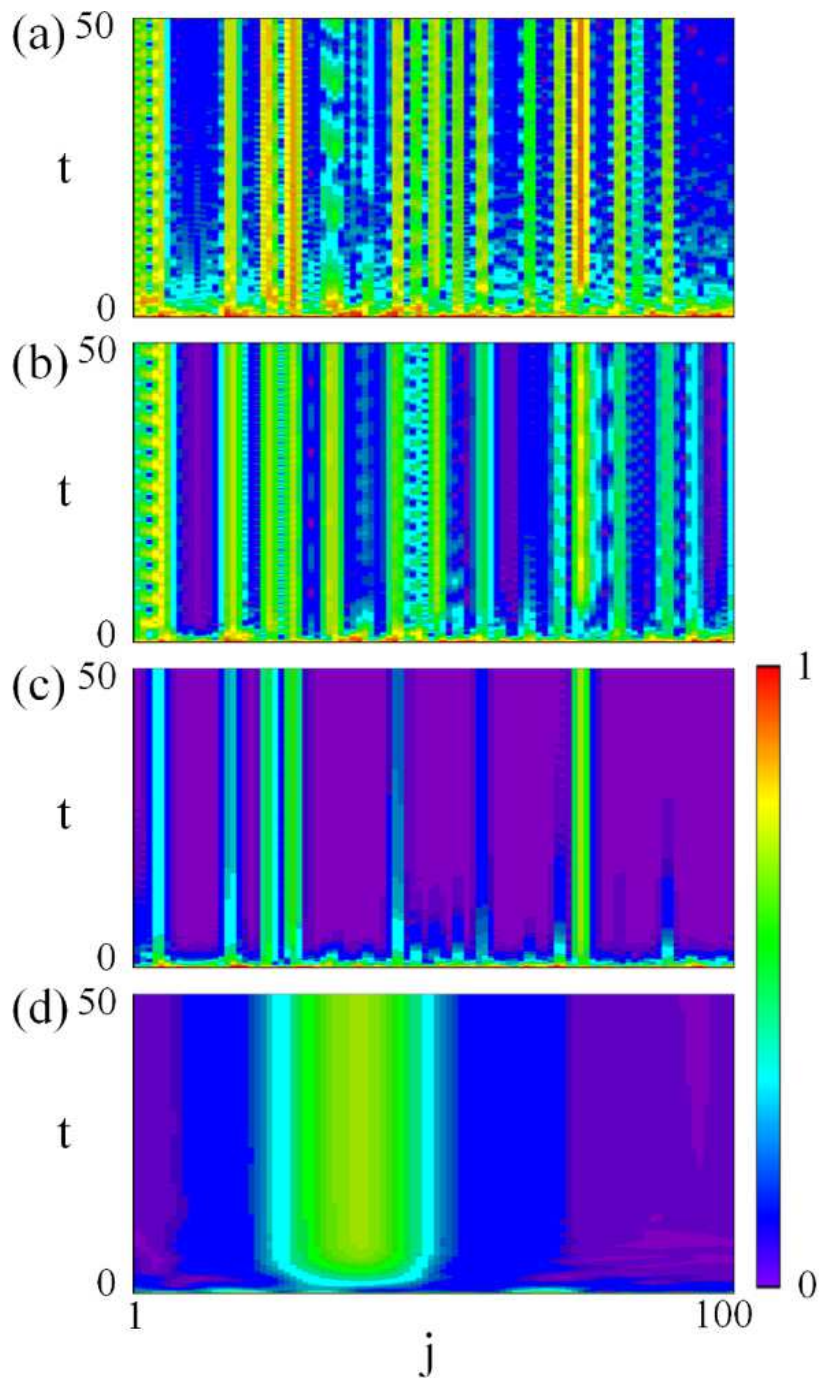}
\caption{(color online).
The spatio-temporal patterns of $|z_j|$ in bidirectional coupling case with ($\omega_{max}$, $k$) = ($10$,$1$), ($10$,$2$), ($20$,$2$), and ($20$,$100$), respectively.
}
\label{fig1}
\end{center}
\end{figure}

\begin{figure}
\begin{center}
\includegraphics[width=\figsizeone\textwidth]{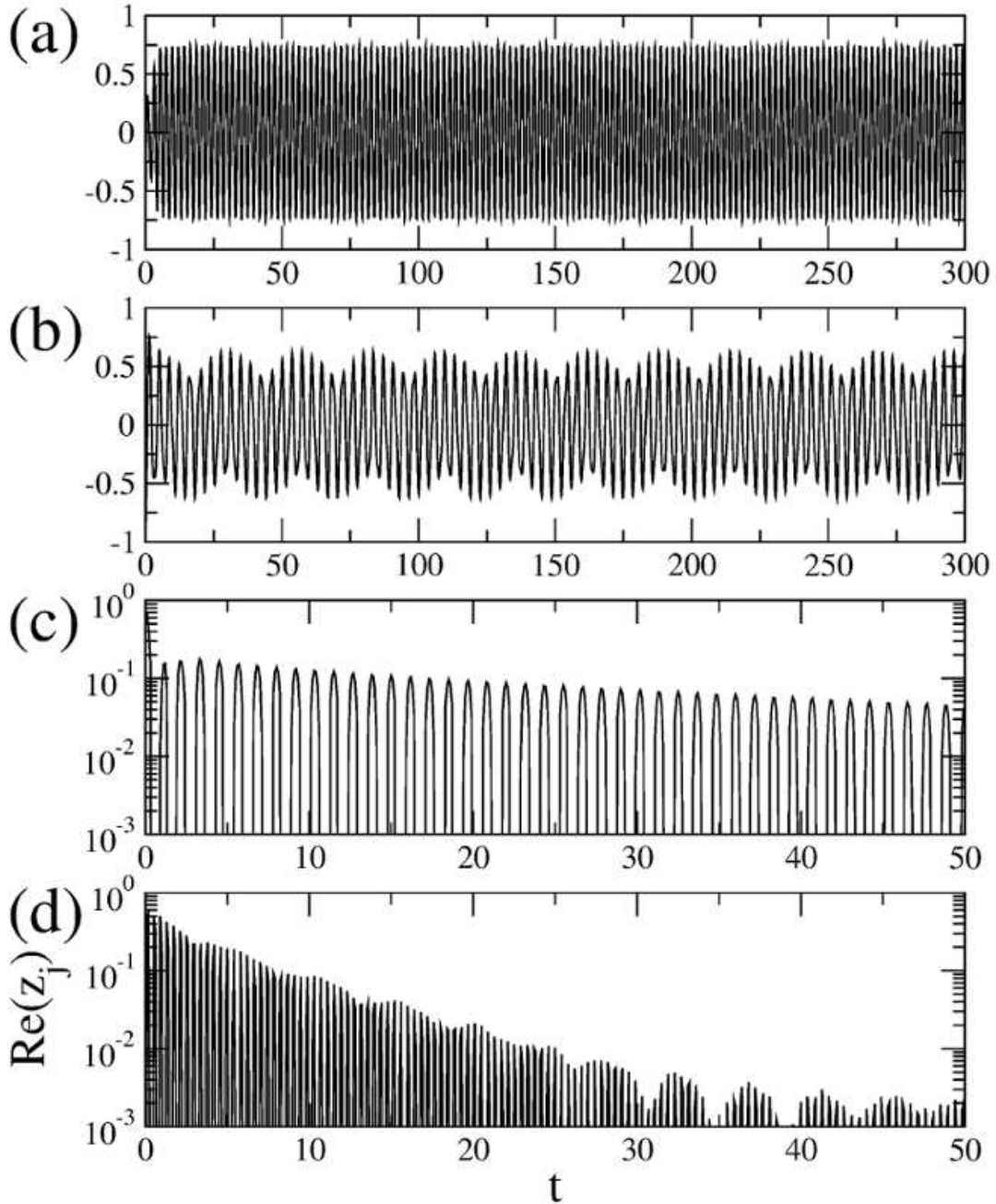}
\caption{Time series of real parts of $z_j$ when (a) $(w_{max},k,j)=(10,2,15)$, (b) $(w_{max},k,j)=(10,2,2)$,
(c) $(w_{max},k,j)=(20,2,58)$, and (d) $(w_{max},k,j)=(20,2,924)$, respectively.	
}
\label{fig2}
\end{center}
\end{figure}

First, we review a ring of coupled Stuart-Landau oscillators with bidirectional coupling.
Figure~\ref{fig1} show the spatio-temporal patterns of amplitudes $\left|z_{j}\right|$ of the first hundred oscillators among a thousand oscillators ($N=1000$) on the plane ($j$, $t$) where $j$ is an index of limit cycle oscillator and $t$ is the time.
We set $\omega_{min}=1$ and change $\omega_{max}$ and $k$.
Stuart-Landau oscillators have same initial amplitudes, $\left|z_j\right| = 1$, and arbitrary initial phases.
Fig.~\ref{fig1} (a) shows the NAD where there is no oscillator showing the amplitude death, $\left|z_j\right|=0$.
There are both oscillators with stationary and oscillatory amplitudes such as Fig.~\ref{fig2} (a) and (b).
Fig.~\ref{fig1} (b), (c), and (d) show the PAD where some oscillators show the time-invariant amplitude death and others do not show that.
Especially, the clustering behaviors appear as shown in Fig.~\ref{fig1} (d), when $k$ is sufficiently large.
In some parameter regions, e.g., ($\omega_{max}$, $k$)=($40$,$5$), all oscillators show the amplitude death, i.e., they show the GAD.

As shown in the spatio-temporal patterns, oscillators have various temporal behaviors and exhibit spatially clustering behaviors in some conditions.
Figure~\ref{fig2} show selected time series of real parts of complex $z_j$, which
exhibit the non-decaying behavior without amplitude oscillations, the non-decaying behavior with amplitude oscillations,
the decaying behavior without amplitude oscillations, and the decaying behavior with amplitude oscillations, respectively.
The spatio-temporal patterns in Fig.~\ref{fig1} show just amplitudes of the time series. Time series without amplitude oscillation in Fig.~\ref{fig2} (a) and (c) make temporally contant patterns in Fig.~\ref{fig1} and time series with amplitude oscillation in Fig.~\ref{fig2} (b) and (d) do oscillatory patterns.
It is noted that there are both temporal behaviors with and without amplitude oscillations in a ring of coupled Stuart-Landau oscillators.
In two coupled dissipative oscillators, the final states of the oscillators are the fixed points if the oscillators exhibit AD.
However, the transient behavior into the AD states shows transitions of freqeuncy locking
as well as amplitude oscillation at exceptional points where there is the balance between coupling strength and difference of
intrinsic angular frequencies of oscillators \cite{Ryu15}.
As a result, the decaying time series without and with amplitude oscillation such as Fig.~\ref{fig2} (c) and (d)
represent the locally narrow and wide distributions of intrinsic angular frequencies, respectively, at fixed coupling strength.

\begin{figure}
\begin{center}
\includegraphics[width=\figsizeone\textwidth]{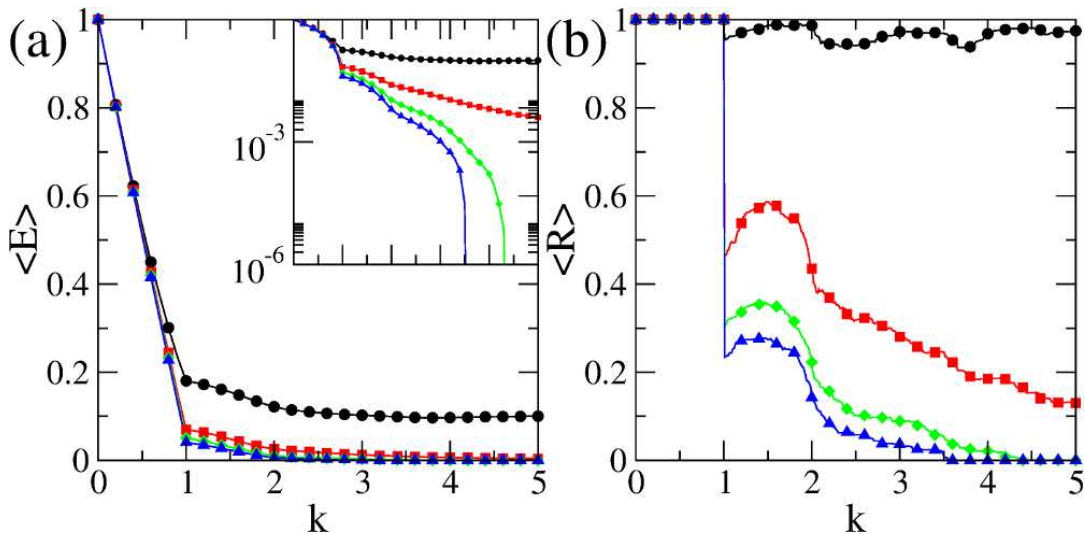}
\caption{(color online).
Time average values of (a) $E(t)$ and (b) $R(t)$ in bidirectional coupling case when $w_{max}=10$ (black circle), $w_{max}=20$ (red rectangle),
$w_{max}=30$ (green diamond), and $w_{max}=40$ (blue triangle), respectively.
Inset shows logarithmic scaled time average values of $E(t)$.
}
\label{fig3}
\end{center}
\end{figure}

In order to classify NAD, PAD, and GAD, we define the incoherent energy $E(t)$
and the normalized number of non-amplitude death sites $R(t)$ as
\begin{eqnarray}
\label{measures}
 E(t) = \frac{\Sigma_{j=1}^N \left|z_j\right|^2}{N}, \\
 R(t) = \frac{N-N_{AD}}{N},
\end{eqnarray}
where $N_{AD}$ is the number of AD sites \cite{Rub00,Yan07}.
The incoherent enregy $E(t)$ gives total sum of intensity of all oscillators without phase information.
It is noted that there is also a corresponding coherent energy which gives total sum of complex variables of all oscillators
including phase information \cite{Rub00}. In our work, they give almost same results.
$R(t)$ is the proportion of oscillators of which amplitude is larger than threshold value for amplitude death, i.e., $\left|z_j\right|>c$.
Figure~\ref{fig3} show the time average of $E(t)$ and $R(t)$, i.e., $\left<E\right>$ and $\left<R\right>$,
when $w_{max}=10$, $20$, $30$, and $40$, respectively.
Here, $\left<E\right>$ and $\left<R\right>$ are calculated after transient time.
NAD, PAD, and GAD can be classified according to $\left<R\right>$.
NAD is for $\left<R\right> =1$, PAD is for $0< \left<R\right> <1$, and GAD is for $\left<R\right> =0$, respectively.
$\left<E\right> >0$ in both NAD and PAD regions and $\left<E\right> =0$ in a GAD region.

The transition for fixed $\omega_{max}$ with increasing $k$ is well investigated in Ref. \cite{Yan07}, here we simply summarize numerical results in order to compare transition of unidirectional case. As the coupling strength $k$ increases from $0$ when $\omega_{max}=10$, $\left<E\right>$ decreases linearly if $k < 1$.
This is the NAD region, i.e., $\left<R\right> =1$.
The $\left<E\right>$ decreases as $k$ increases beyond $1$ but the behavior is not linear.
The $\left<E\right>$ does not decrease further when $k \gtrsim 3.7$, and even increases for higher value of $k$ (which is not shown here).
$\left<R\right> < 1$ represents the PAD region when $k >1$ but the PAD regions change into the NAD regions ($\left<R\right>=1$) if $k$ is sufficiently large.
When $\omega_{max}=10$, $\left<E\right> > 0$ and $\left<R\right> > 0$ for all $k$, that is, there is no GAD region.
When $\omega_{max}=20$, there are also NAD region if $k<1$ and PAD region if $k>1$ but the GAD region does not exist.
The $\left<E\right>$ increases again as $k$ increases beyond about $7.6$.
As a result, there are no GAD regions when $\omega_{max}=10$ and $20$.
When $\omega_{max}=30$ and $40$, $\left<E\right>$ decreases linearly and $\left<R\right>=1$ if $k<1$, which represent NAD regions.
$\left<E\right>$ decreases in more complex manner and $0<\left<R\right><1$ if $k>1$, corresponding to PAD regions.
Finally, both $\left<E\right>$ and $\left<R\right>$ become zero if $k \gtrsim 4.4$ and $3.6$, respectively, i.e., GAD regions.

The bidirectional coupling term of Eq.~(\ref{slosc}) can be divided into two parts, $-k z_{j}$ and $k (z_{j-1} + z_{j+1}) / 2$.
Due to the first term, when $k<1$, $\left<E\right>$ decreases monotonously as $k$ increases because the dynamical behavior of individual oscillator is dominant. However, as $k$ increases more, due to the second term, the instability originating from couplings between oscillators increases and then the oscillations can revive becasue the network strucuture is more dominant than the dynamcal behavior of individual oscillators.
We will discuss this again in Sec. IV in terms of the eigenvalue analysis of Jacobian matrix at the origin.

\subsection{Unidirectional coupling case}

\begin{figure}
\begin{center}
\includegraphics[width=\figsizetwo\textwidth]{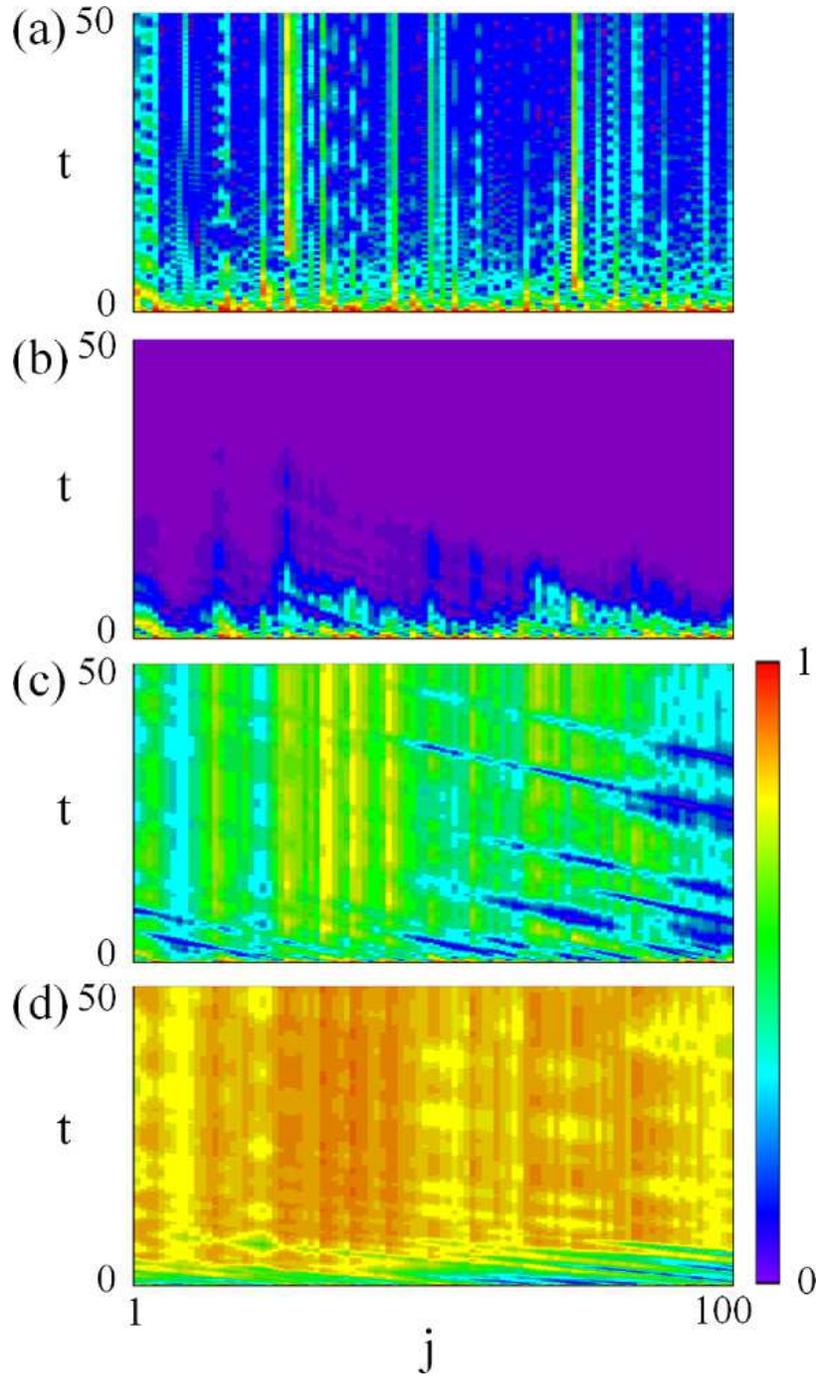}
\caption{(color online).
The spatio-temporal patterns of $|z_j|$ in unidirectional coupling case with ($\omega_{max}$, $k$) = ($10$,$1$), ($10$,$2$), ($10$,$4$), and ($10$,$10$), respectively.
}
\label{fig4}
\end{center}
\end{figure}

Next, we consider a ring of coupled Stuart-Landau oscillators with unidirectional coupling.
Figure~\ref{fig4} show the spatio-temporal patterns of $\left|z_j\right|$ on the plane ($j$, $t$).
We set $w_{min}=1$ and change $w_{max}$ and $k$.
Fig.~\ref{fig4} (a) shows the NAD where there is no oscillator showing AD.
Fig.~\ref{fig4} (b) shows the GAD, i.e., all oscillators cease to oscillate after transient times.
Fig.~\ref{fig4} (c) and (d) show different NAD.
The spatio-temporal patterns are totally different from those of bidirectional coupling case as follows:
(i) There is no PAD region in the unidirectional coupling case.
If we set $z_{j+1}=0$ of Eq.~(\ref{slosc}), the equation can be considered as uncoupled oscillators,
$ \dot{z}_j=(1-k +i \omega_j - |z_j|^2) z_j$, in which the stability of an oscillator are determined by sign of $(1-k)$.
When $k$ is larger than $1$ in the unidirectional coupling case, all oscillators should exhibit the AD if any one oscillator
shows the AD, i.e. there are no PAD regions.
(ii) There are oblique line structures in the spatio-temporal patterns, which represent the unidirectionality of signal flow.
The oblique lines originate from the relation between $z_j (t+\Delta t)$ and $z_{j+1} (t)$ in Eq.~(\ref{slosc}), in other words, the signal of $z_{j+1}$ arrive at $z_{j}$ after time $\Delta t$. Considering only two terms $\dot{z_j}$ and $F(z_{j+1} - z_{j})$, Eq.~(\ref{slosc}) can be rewritten as
\begin{equation}
\frac{\Delta z_j (t)}{\Delta t} = -k \frac{\Delta z_j (t)}{\Delta j}.
\end{equation}
Finally, we obtain the relation, $\Delta t/\Delta j=-1/k$, which corresponds to the slope of the oblique lines.
(iii) There is no local clustering behaviors even if $k$ is sufficiently large.
The local clustering in a ring of coupled oscillators is related to the enhancement of coherence between oscillators. In unidirectional coupling case, however, the coherence cannot be enhanced because there is no recurrence when the system is sufficiently large.

\begin{figure}
\begin{center}
\includegraphics[width=\figsizeone\textwidth]{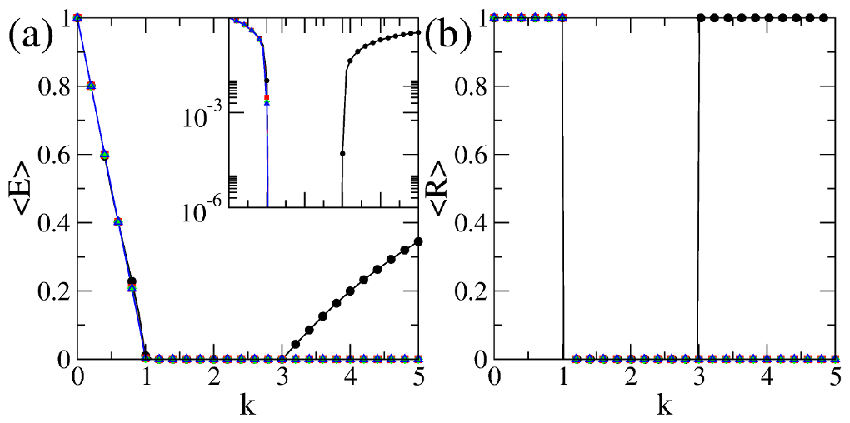}
\caption{(color online).
Time average values of (a) $E(t)$ and (b) $R(t)$ in unidirectional coupling case when $w_{max}=10$ (black circle), $w_{max}=20$ (red rectangle),
$w_{max}=30$ (green diamond), and $w_{max}=40$ (blue triangle), respectively.
Inset shows logarithmic scaled time average values of $E(t)$.
}
\label{fig5}
\end{center}
\end{figure}

Figure~\ref{fig5} show $\left<E\right>$ and $\left<R\right>$ when $w_{max}=10$, $20$, $30$, and $40$, respectively.
When $\omega_{max}=10$, there are NAD regions if $k < 1$ or $k > 3$ and GAD regions if $1 < k < 3$ in Fig.~\ref{fig5} (a).
When $\omega_{max}=20$, there are also NAD regions if $k < 1$ or $k \gtrsim 20$ and GAD regions if $1 < k \lesssim 20$ in Fig.~\ref{fig5} (b).
In the unidirectional coupling cases, there are no qualitative changes of $\left<E\right>$ and $\left<R\right>$ as varying the value of $\omega_{max}$.
The difference in $\omega_{max}$ gives only different boundary of GAD region.

\section{Eigenvalue analysis}

In this section, we analyze the eigenvalues of Jacobian matrix at the origin.
The real and imaginary parts of eigenvalues of the Jacobian matrix corresponds to the decay (or growing) rates and the angular frequency of the orbit near the origin, respectively.
The stability of GAD is determined by the sign of maximal value of real parts \cite{Ryu15}
and the distribution of imaginary parts are related to the coherence of oscillators.

\subsection{Jacobian matrix at origin}

\begin{figure}
\begin{center}
\includegraphics[width=\figsizeone\textwidth]{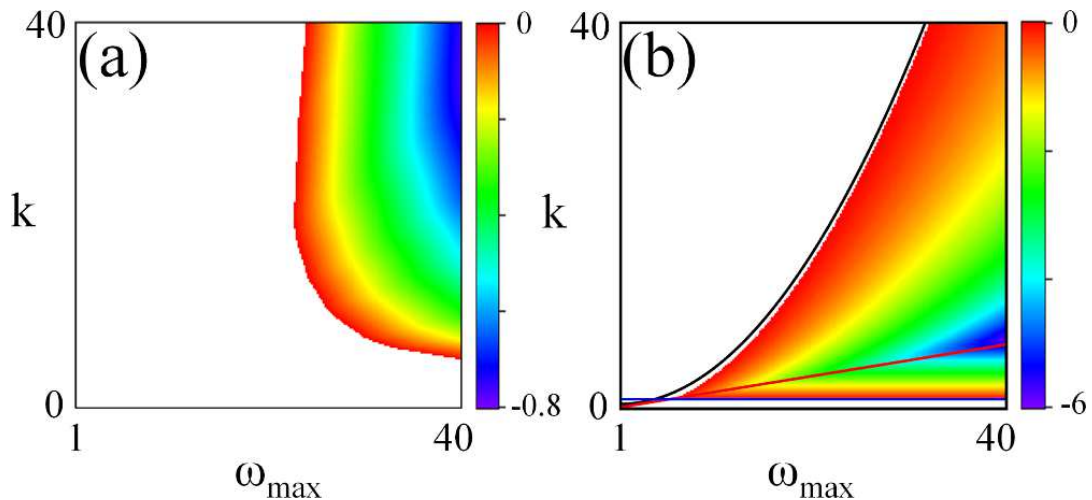}
\caption{(color online). Maximal values of the real parts of eigenvalues of linearized Jacobian matrix at origin
in (a) bidirectional and (b) unidirectional coupling cases, respectively.
The stability diagrams are obtained by ensemble average of $10$ random sets.
The colored and white regions represent negative and positive values, respectively.
Colored regions are GAD regions where maximal real parts of eigenvalues are smaller than zero.
From top to bottom, the black curve, red, and blue straight lines correspond to $k=(1+W_{\omega}^2/12)/2$, $k=W_{\omega}/6$, and $k=1$, respectively.
}
\label{fig6}
\end{center}
\end{figure}

\begin{figure}
\begin{center}
\includegraphics[width=\figsizeone\textwidth]{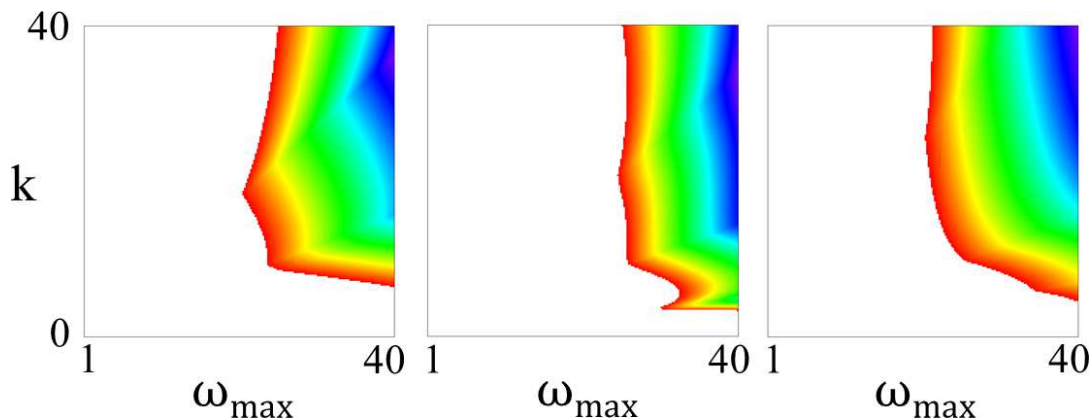}
\caption{(color online). Three examples of the stability diagrams for the bidirectional coupling case with different random sets. The colored regions represent GAD regions. Color scales are same as those of Fig.~\ref{fig6} (a).
}
\label{fig6b}
\end{center}
\end{figure}

For the bidirectional coupling case, the Jacobian matrix at origin is
\begin{equation}
 \label{jac_bi}
M=\left(\begin{array}{ccccc}
 A_1 & \frac{k}{2}I & \cdots & 0 & \frac{k}{2}I \\
 \frac{k}{2}I & A_2 & \cdots & 0 & 0 \\
 \vdots & \vdots & \ddots & \vdots & \vdots \\
 0 & 0 & \cdots & A_{N-1} & \frac{k}{2}I \\
\frac{k}{2}I & 0 & \cdots & \frac{k}{2}I & A_N
\end{array}\right),
\end{equation}
where
\begin{equation}
A_{j}=\left(\begin{array}{cc}
1-k & -\omega_{j}\\
\omega_{j} & 1-k
\end{array}\right)
~\mathrm{and}~
I=\left(\begin{array}{cc}
1 & 0\\
0 & 1
\end{array}\right).
\end{equation}
For the unidirectional coupling case, the Jacobian matrix at origin is
\begin{equation}
 \label{jac_uni}
M=\left(\begin{array}{ccccc}
 A_1 & k I & \cdots & 0 & 0 \\
 0 & A_2 & \cdots & 0 & 0 \\
 \vdots & \vdots & \ddots & \vdots & \vdots \\
 0 & 0 & \cdots & A_{N-1} & k I \\
k I & 0 & \cdots & 0 & A_N
\end{array}\right).
\end{equation}
The condition of GAD is that the maximal real parts of eigenvalues are smaller than zero.
From this condition, we can obtain GAD region on the plane ($\omega_{max}$, $k$).
Figure~\ref{fig6} (a) and (b) show the stability diagrams which are obtained by the ensemble averages of $10$ different random sets of natural frequencies in the bidirectional and unidirectional cases, respectively.
Besides the boundary shapes, two stability diagrams are totally different.
The stability diagrams for the bidirectional coupling case are different if the random sets are different as shown in Fig.~\ref{fig6b}.
However, the stability diagram for the unidirectional coupling case is almost independent of random set.
That is, even if the random set is changed, the stability diagrams are same as those of Fig.~\ref{fig6} (b).
In addition, the results in the unidirectional coupling case are very similar to the results in two coupled Stuart-Landau oscillators \cite{Ryu15}.
In two coupled Stuart-Landau oscillators, the stability regions have two boundary lines,
$k=1$ and $k=(1+\Delta\omega^2 /4)/2$ where $\Delta\omega=\omega_2 - \omega_1$.
There is also a local mininum line relating to the exceptional point, $k=\Delta\omega/2$. 
In a ring of coupled Stuart-Landau oscillators, the average absolute difference of intrinsic frequencies of two oscillators is
$\overline{\Delta\omega} = W_\omega/3$, where $W_\omega=\omega_{max}-\omega_{min}$,
if the number of oscillators are sufficiently large.
In Fig.~\ref{fig6} (b), the stability regions have two boundary lines, $k=1$ and $k=(1+{W_{\omega}}^2 /12)/2$,
and one local minimum line, $k=\overline{\Delta\omega}/2 = W_\omega/6$.

The unidirectional coupling has distinct advantages over bidirectional coupling to achieve the GAD which can be considered as the control or stabilization of oscillatory systems. First, a wide distribution of natural frequency is needed to get GAD region in the case of bidirectional coupling. This corresponds to the large minimum value of $\omega_{max}$ in Fig.~\ref{fig6} (a). In the case of unidirectional coupling, however, small $\omega_{max}$ is sufficient to get GAD region as shown in Fig.~\ref{fig6} (b). This means that the system can be easily stabilized by weak disorder in the case of unidirectional coupling. The threshold of coupling strength $k$ in the case of unidirectional coupling is also smaller than that in the case of bidirectional coupling. Next, while there are PAD region between NAD and GAD regions in the case of bidirectional coupling, NAD region directly changes into GAD region in the case of unidirectional coupling. In addition, while GAD regions in the case of bidirectional coupling are different accroding to the random ensemble of $\omega_{j}$ as shown in Fig.~\ref{fig6b}, GAD regions in the case of unidirectional coupling do not change even if different random ensemble of $\omega_j$ is used. As a result, we can anticipate the condition for GAD in the case of unidirectional coupling, irrespective of random ensemble of $\omega_j$

\begin{figure}
\begin{center}
\includegraphics[width=\figsizeone\textwidth]{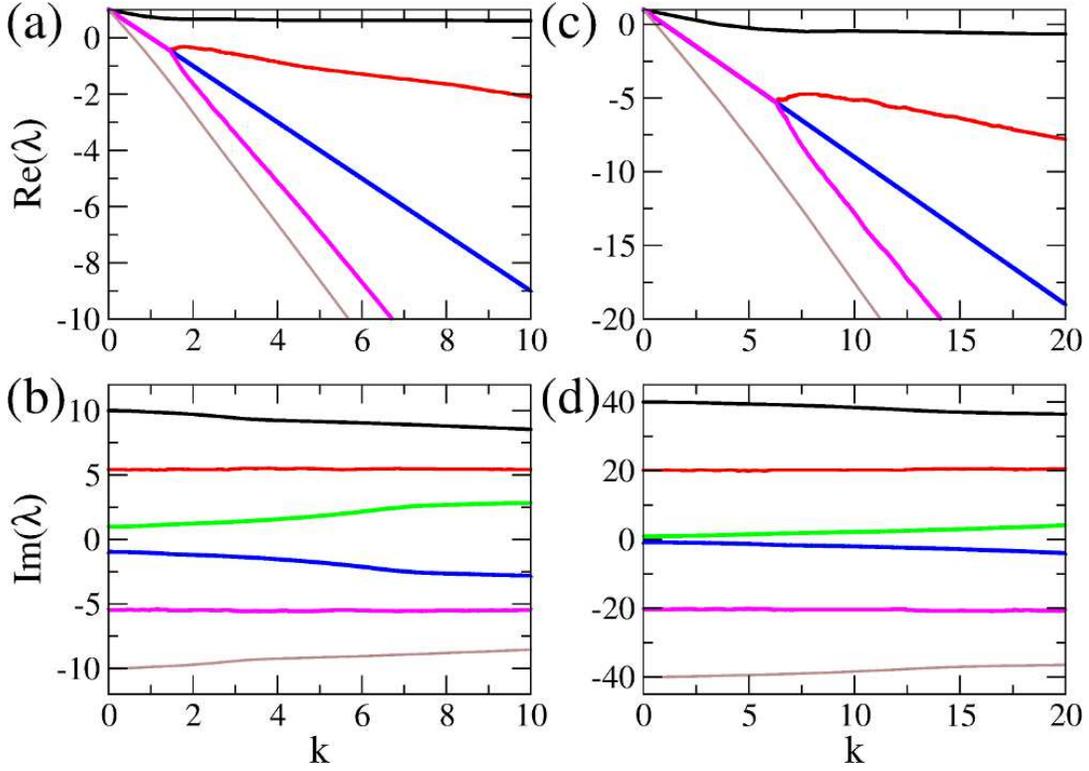}
\caption{(color online).
(a) Real and (b) imaginary parts of eigenvalues in bidirectional coupling case with $\omega_{max}=10$.
(c) Real and (d) imaginary parts of eigenvalues in bidirectional coupling case with $\omega_{max}=40$.
Black circle, red rectangle, green diamond, blue triangle up, magenta triangle down, and brown star represent
$1$st, $500$th, $1000$th, $1001$th, $1501$th, and $2000$th large real and large imaginary parts of eigenvalues, respectively.
}
\label{fig7}
\end{center}
\end{figure}

\begin{figure}[b!]
\begin{center}
\includegraphics[width=\figsizeone\textwidth]{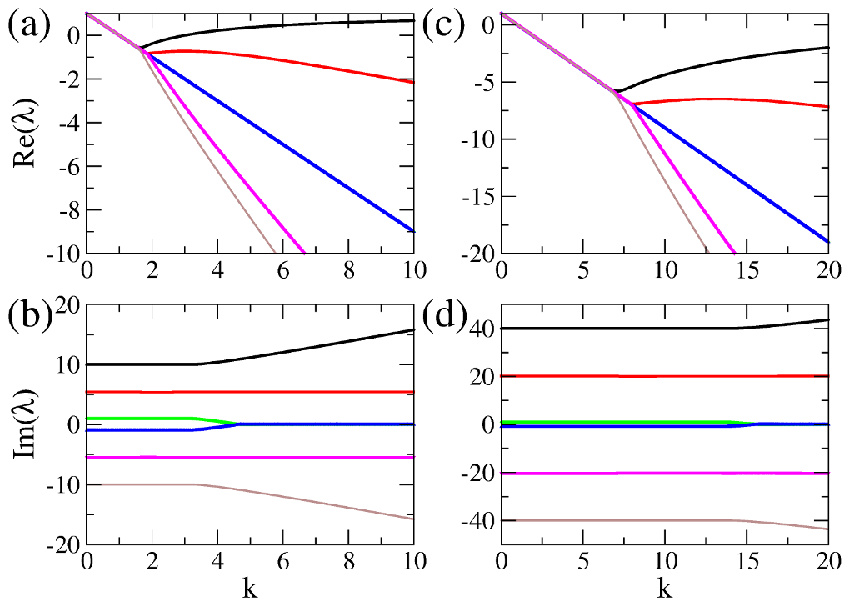}
\caption{(color online).
(a) Real and (b) imaginary parts of eigenvalues in unidirectional coupling case with $\omega_{max}=10$.
(c) Real and (d) imaginary parts of eigenvalues in unidirectional coupling case with $\omega_{max}=40$.
Black circle, red rectangle, green diamond, blue triangle up, magenta triangle down, and brown star represent
$1$st, $500$th, $1000$th, $1001$th, $1501$th, and $2000$th large real and large imaginary parts of eigenvalues, respectively.
}
\label{fig8}
\end{center}
\end{figure}

Figure~\ref{fig7} and \ref{fig8} show selected real and imaginary parts of eigenvalues in the bidirectional and unidirectional coupling cases as a function of $k$.
Two middle values of real parts, i.e., $1000$th and $1001$th largest real parts, show the relation $\mathrm{Re}(\lambda)=1-k$,
which corresponds to the linear decreasing of $\left<E\right>$ when $k<1$ in Fig.~\ref{fig3} and \ref{fig5}.
Considering the unidirectional coupling case in Fig.~\ref{fig8}, maximal real parts have branching points near $k \sim W_\omega/6$ such as Fig.~\ref{fig6} (b).
All real parts of complex eigenvalues are $\mathrm{Re}(\lambda)=1-k$ before first branching point at $k \sim W_\omega /6$
and the distribution of real parts also becomes wider after first branching point.
We note that this is a general property of unidirectional coupling case if the system size is sufficiently large.
The tendency of maximal real parts of complex eigenvalues are changed to increasing from decreasing at a branching point as $k$ increases,
which is important to reviving oscillations when the coupling strength is large.
These branching points also relate to the enhancement of coherence of oscillators
because two corresponding imaginary parts merge into one value when the one real part splits into two values via exceptional points \cite{Ryu15}.
In the bidirectional coupling cases of Fig.~\ref{fig7}, the maximal real parts show more complex behaviors without clear first branching point,
which make different stability diagrams according to the different random sets.

The imaginary parts of eigenvalues show two different regions in Fig.~\ref{fig8} (b) and (d).
The imaginary parts of eigenvalues do not vary if $k$ is small but the width of imaginary parts becomes wider as $k$ increases
because the width of imaginary parts of complex eigenvalues of the circulant matrix for the unidirectional coupling case increases (see Fig.~\ref{fig10} (d)).
In the bidirectional coupling cases of Fig.~\ref{fig7} (b) and (d), the width of imaginary parts of eigenvalues become narrower as $k$ increases
because imaginary parts of complex eigenvalues of the block circulant matrix for bidirectional coupling case are constant (see Fig.~\ref{fig9} (d)).

\subsection{Dynamical behavior vs. network structure}

Let's rewrite the Jacobian matrix in the bidirectional coupling case of Eq.~(\ref{jac_bi}) as 
\begin{widetext}
\begin{eqnarray}
 \label{jac2_bi}
M &=& M_0+ \frac{W_{w}}{2} M_1 + k M_2 \\\nonumber
  &=&
\left(\begin{array}{ccccc}
C & 0 & \cdots & 0 & 0\\
0 & C & \cdots & 0 & 0 \\
\vdots & \vdots & \ddots & \vdots & \vdots \\
0 & 0 & \cdots & C & 0 \\
0 & 0 & \cdots & 0 & C
\end{array}\right)+\frac{W_{\omega}}{2}\left(
\begin{array}{ccccc}
D_1 & 0 & \cdots & 0 & 0\\
0 & D_2 & \cdots & 0 & 0 \\
\vdots & \vdots & \ddots & \vdots & \vdots \\
0 & 0 & \cdots & D_{N-1} & 0 \\
0 & 0 & \cdots & 0 & D_N
\end{array}\right)+k\left(
\begin{array}{ccccc}
-I & \frac{1}{2}I & \cdots & 0 & \frac{1}{2}I \\
\frac{1}{2}I & -I & \cdots & 0 & 0 \\
\vdots & \vdots & \ddots & \vdots & \vdots \\
0 & 0 & \cdots & -I & \frac{1}{2}I \\
\frac{1}{2}I & 0 & \cdots & \frac{1}{2}I & -I
\end{array}\right),
\end{eqnarray}
\end{widetext}
where
\begin{equation}
C=\left(\begin{array}{cc}
1 & -\bar{\omega}\\
\bar{\omega} & 1
\end{array}\right)
~\mathrm{and}~
D_{j}=\left(\begin{array}{cc}
0 & -\xi_{j}\\
\xi_{j} & 0
\end{array}\right).
\end{equation}
$\bar{\omega}=\frac{\Sigma_{j=1}^{N} \omega_j}{N}$, $\omega_j - \bar{\omega} = \xi_j W_\omega /2$,
and $\xi_j$ is a random number between $-1$ and $1$.
A matrix $M_0$ represents the original dynamics of Stuart-Landau oscillators with same angular frequency $\bar{\omega}$ and
a matrix $M_1$ represents the applied disorder term of angular frequency.
As $W_\omega$ increases, the disorders of systems increase.
The matrix $M_2$ is a coupling term originating from ring structures, which is a kind of circulant matrices.

\begin{figure}
\begin{center}
\includegraphics[width=\figsizeone\textwidth]{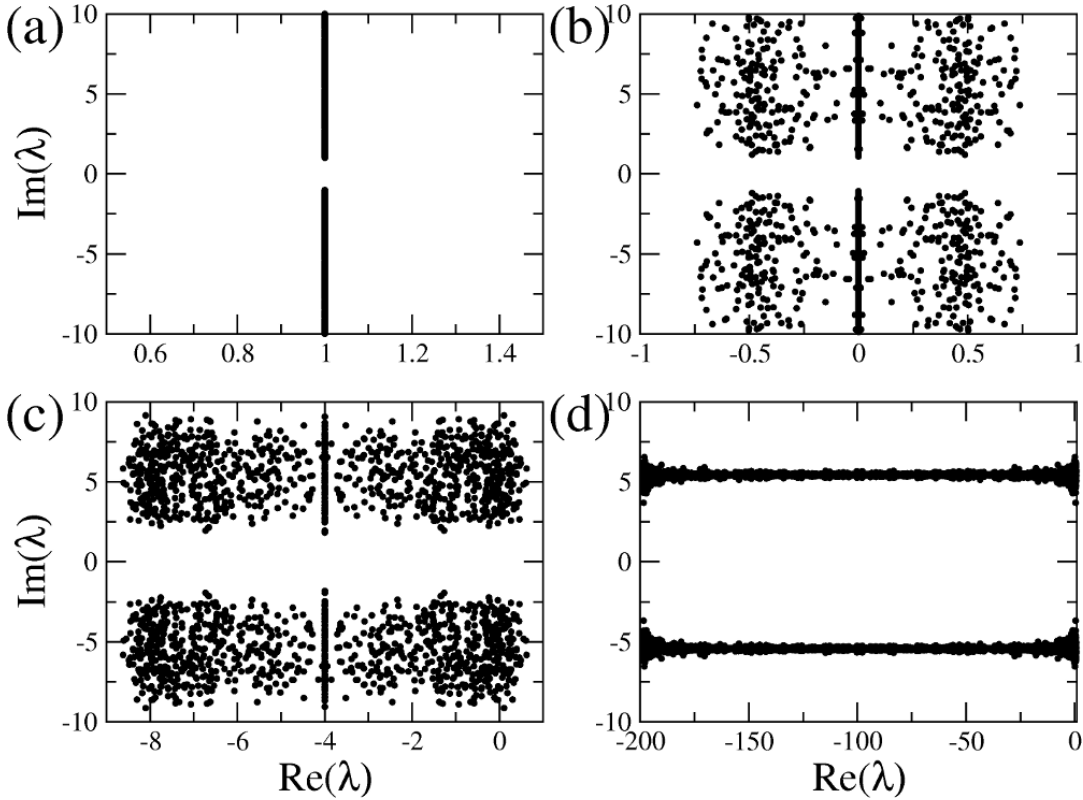}
\caption{
Complex eigenvalues in bidirectional coupling case with $\omega_{max}=10$ when (a) $k=0$, (b) $k=1$, (c) $k=5$, and (d) $k=100$, respectively.
}
\label{fig9}
\end{center}
\end{figure}

Figure~\ref{fig9} show the complex eigenvalues in the bidirectional coupling case. First we consider non-coupled Stuart-Landau oscillators ($k=0$).
The complex eigenvalues are a vertical line on a complex plane because the system has randomly distributed angular frequencies.
The distribution of imaginary parts of eigenvalues equal to that of $\omega_j$ of our systems.
As $k$ increases, two changes appear due to $M_2$. (i) The mean value of real parts is inversely proportional to $k$ because of negative diagonal elements of $M_2$.
This correponds to the decreasing behaviors of mean values of real parts of eigenvalues as shown in Fig.~\ref{fig7} (a) and (c).
(ii) The distributions of real parts become wider due to off-diagonal elements of $M_2$.
Finally, if $k$ is sufficiently large, the complex eigenvalues form two symmetric horizontal lines
which are the complex eigenvalues of symmetric circulant matrix such as $M_2$.
Two symmetric horizontal lines of complex eigenvalues mean strong coherence of oscillators
because the distribution of imaginary parts correspond to that of angular frequencies of oscillators.
The clustering behaviors in the case of bidirectional coupling as shown in Fig.~\ref{fig1} (d) can be understood in terms of the narrower distribution of the imaginary parts of eigenvalues, which implies easier synchronization between oscillators. As discussed in Ref. \cite{Yan07}, synchronization prevents amplitude death, therefore narrower distribution of imaginary parts of eigenvalues results in clustered activation against global amplitude death.

\begin{figure}
\begin{center}
\includegraphics[width=\figsizeone\textwidth]{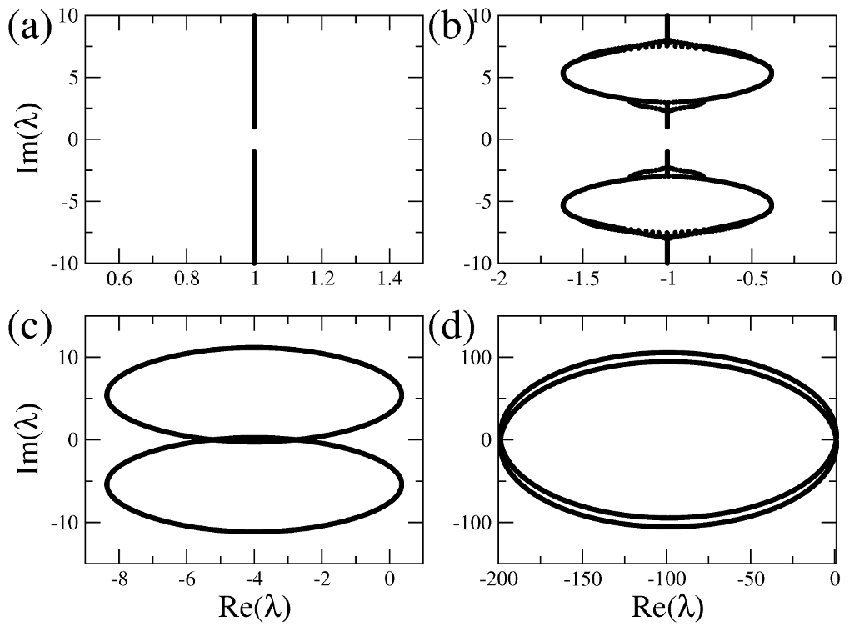}
\caption{
Complex eigenvalues in unidirectional coupling case with $\omega_{max}=10$ when (a) $k=0$, (b) $k=2$, (c) $k=5$, and (d) $k=100$, respectively.
}
\label{fig10}
\end{center}
\end{figure}

For the case of unidirecitional coupling, the Jacobian matrix at the origin can be transformed into
$M=M_{0} +\frac{W_{\omega}}{2}M_{1} + k M_{3}$ where $M_{0}$ and $M_{1}$ are same for Eqs.~(\ref{jac2_bi}).
Here, $M_{3}$ is represented as
\begin{equation}
M_3=
\left(
\begin{array}{ccccc}
-I & I & \cdots & 0 & 0 \\
0 & -I & \cdots & 0 & 0 \\
\vdots & \vdots & \ddots & \vdots & \vdots \\
0 & 0 & \cdots & -I & I \\
I & 0 & \cdots & 0 & -I
\end{array}\right)
\label{m3}
\end{equation}
Figure~\ref{fig10} show the complex eigenvalues in the unidirectional coupling cases.
Contrary to the bidirectional coupling case, if $k$ is sufficiently large,
the complex eigenvalues form two circles which correspond to the complex eigenvalues of asymmetric circulant matrix such as $M_3$.
The circularly distributed complex eigenvalues correspond to the results in the non-Hermitian random matrices with directed hoppings,
which arise in the physics of randomly pinned superconducting vortex lines \cite{Hat96, Hat97} and 
in biological networks \cite{Ami16}.
There is no clustering behavior in the case of unidirectional coupling as shown in Fig.~\ref{fig4}
because of the wide distribution of the imaginary parts of eigenvalues.

\section{Discussion and Summary}

In this work, the intrinsic angular frequencies $\omega_j$ of uncoupled $j$th limit cycle oscillator is a random number between $\omega_{min}$ and $\omega_{max}$ with uniform distribution. If we use the $\omega_j$ with different distribution other than uniform, our results will be different because the stability of AD state is totally determined by the difference between frequencies of nearest neighbors, which is closely related to the distribution of $\omega_j$. For instance, considering Gaussian distribution of $\omega_j$ which has higher probability near the average $\omega_j$, it is more difficult to get AD state because of smaller average difference between frequencies of nearest neighbors.

In the bidirectional coupling case, the periodic boundary condition is not essential for our results
if the number of oscillators is sufficiently large.
However, the periodic boundary condition has a very important role in the amplitude death phenomena in the unidirectional coupling case.
For an extreme example, if we consider no boundary condition, there is no amplitude death because there is an unperturbed $N$-th oscillator which always oscillates and therefore preventes the amplitude death of oscillators.

We have studied the amplitude death in a ring of coupled nonlinear oscillators with both bidirectional
and cyclically unidirectional coupling, of which natural frequencies are distributed in a random way.
We have found the amplitude death phenomena in both cases of bidirectional and unidirectional couplings
and discussed the differences between two coupling cases.
There are three main differences; there exists neither partial amplitude death nor local clustering behavior
but oblique line structure which represents directional signal flow on the spatio-temporal patterns in the unidirectional coupling case.
The unidirectional coupling has the advantage over the bidirectional coupling to achieve the global amplitude death with small coupling strength and weak disorder in a ring of coupled oscillators.
Finally, we have explained the results using the eigenvalue analysis of Jacobian matrix at the origin and
also discussed the transition of dynamical behavior coming from connection structure as coupling strength increases.

\section*{Acknowledgments}
This research was supported by Project Code (IBS-R024-D1).
This research was supported by National Institute for Mathematical Sciences (NIMS) funded by the Ministry of Science, ICT \& Future Planning (A22200000).
This research was supported by the National Research Council of Science and Technology (NST) grant by the Korea government (MSIP) (No. CRC-16-01-KRICT).

\end{document}